\documentclass[twocolumn,amsmath,amssymb,prl,showpacs]{revtex4}

\usepackage{graphicx}% Include figure files
\usepackage{dcolumn}% Align table columns on decimal point
\usepackage{bm}% bold math

\begin{document}

\title{Anomalous-Nernst and anisotropic magnetoresistive heating in a lateral spin valve}

\author{A. Slachter}
 \email{A.Slachter@gmail.com}
\author{F.L. Bakker}
\author{B. J. van Wees}%

\affiliation{Physics of Nanodevices, Zernike Institute for Advanced Materials, University
of Groningen, The Netherlands}

\date{\today}

\begin{abstract}
We measured the anomalous-Nernst effect and anisotropic magnetoresistive heating in a lateral multiterminal Permalloy/Copper spin valve using all-electrical lock-in measurements. To interpret the results, a three-dimensional thermoelectric finite-element-model is developed. Using this model, we extract the heat profile which we use to determine the anomalous Nernst coefficient of Permalloy R$_{N}$=0.13 and also determine the maximum angle $\theta=8^{\circ}$ of the magnetization prior to the switching process when an opposing non-collinear 10$^{\circ}$ magnetic field is applied.
\end{abstract}

\pacs{72.15.Jf, 72.25.-b, 85.75.-d, 85.80.-b}

\maketitle

The connection between thermoelectricity and spintronics\cite{Zutic} has recently attracted a lot of attention\cite{Saitoh,Slachter} which led to the subfield called spin-caloritronics\cite{BauerSpinCaloritronics}. Although thermoelectric effects are typically regarded small, we have recently shown that they can be dominant in lateral multiterminal devices such as the non-local spin valve\cite{Bakker,Slachter}. Here we demonstrate two thermal effects which can accompany such new functionality in nanoscale spin-caloritronic devices: the anomalous-Nernst effect and anisotropic magnetoresistive heating. We show that both effects can dominate the thermoelectric behavior and can be modeled accurately.

The anomalous-Nernst effect can be interpreted as the thermoelectric equivalent of the anomalous-Hall effect\cite{Nagaosa,Behnia}. When a temperature gradient is applied to a ferromagnet, a voltage gradient perpendicular to the plane made by the magnetization and temperature gradient develops and vice versa. Both effects are related to each other and are described by the same Nernst coefficient R$_{N}$. The first effect is governed by the following equation:

\begin{equation}\label{eq:AnNernst}
\vec \nabla V_{N}= - S_{N}  \vec m \times \vec\nabla T
\end{equation}

\noindent here $\vec m$ is the unit vector pointing in the magnetization direction, T the temperature and $\vec\nabla$V$_{N}$ the resulting voltage gradient due to anomalous-Nernst effect. S$_{N}$=R$_{N}$S is the transverse Seebeck coefficient representing the strength of the effect, which is a fraction of the Seebeck coefficient S.

\begin{figure}[b]
\includegraphics[width=8.8cm,keepaspectratio=true]{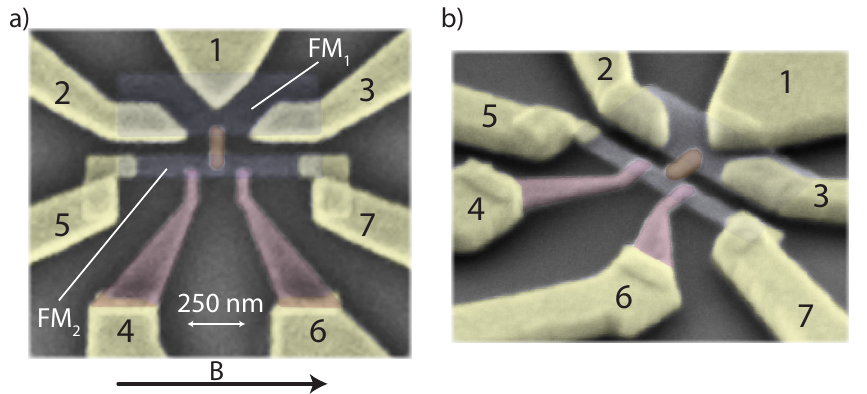}
\caption{\label{fig:1} (color) Colored Scanning Electron Microscope images of the fabricated device. a) Top view of the device. The two ferromagnets (blue) are connected by a copper strip (brown). FM$_{1}$ is connected by three thick gold heat sinks (yellow) through which we can send a charge current to heat it. FM$_{2}$ is also connected by two gold heat sinks (yellow) but have two additional NiChrome contacts (pink). The magnetizations $\vec M_{1}$ and $\vec M_{2}$ are selectively switched by applying an opposing magnetic field $\vec B$. b) Three dimensional image of the device illustrating the thick gold contact used as thermal heat sinks.}
\end{figure}

The anisotropic magnetoresistance (AMR) describes how the resistance of a ferromagnet changes with respect to the angle $\theta$ between the magnetization and the current direction. The conductivity of the ferromagnet is given by $\sigma_{FM}$=$\sigma_{\parallel}$(1+R$_{AMR}(\cos^{2}(\theta)-1))$ where $\sigma_{\|}$ is the conductivity measured when the direction of the current is parallel to the magnetization and R$_{AMR}$ a small fraction. When a current is sent through a ferromagnet, the Joule heating of this ferromagnet depends on the resistance of the magnet. Therefore, the Joule heating of a ferromagnet depends on the angle between the magnetization and the direction of the current. Because the non-local voltages measured in lateral multiterminal device depend on the generated heat\cite{Bakker}, this angle can be deduced from measurements. We refer to this effect as anisotropic magnetoresistive heating.

\begin{figure}[!t]
\includegraphics[width=8.8cm,keepaspectratio=true]{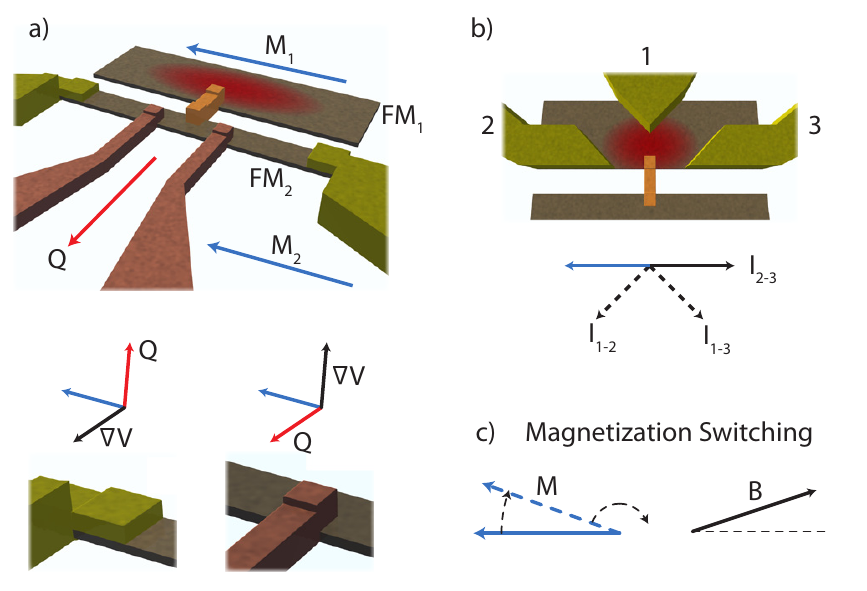}
\caption{\label{fig:2} (color online) Illustration of the anomalous-Nernst effect and Magnetoresistive heating. a) The Joule heating of FM$_{1}$ induces a heat flow Q through FM$_{2}$ and the four contacts connecting it. The anomalous-Nernst effect induces voltage gradients in the ferromagnet perpendicular to the heat flow and magnetization M$_{2}$. b) Three contacts are present on FM$_{1}$ to send the current parallel or under a $\pm$45 degree angle with respect to the magnetization of FM$_{1}$. c) When the opposing magnetic field has a small angle with respect to the antiparallel direction of the magnetization, the magnetization first rotates prior to switching at its switching field increasing or decreasing the Joule heating depending on the orientation of the current.}
\end{figure}

To demonstrate both effects, we fabricated a multiterminal lateral spin valve. This device is shown in figure 1. It consists of two Permalloy (Ni$_{80}$Fe$_{20}$) ferromagnets connected by a highly conductive copper strip. The first ferromagnet FM$_{1}$ is provided with three thick highly thermally conductive Ti/Au contacts which allows to locally heat this ferromagnet by sending currents through it. The generated heat is transported to the second ferromagnet FM$_{2}$ by the thermally conductive copper strip. This heat can be detected by measuring the temperature of this ferromagnet close to the Py/Cu interface. We do this by providing two thermocouples to FM$_{2}$. The outer sides are thermally anchored by two gold contacts, while close to the interface two NiChrome (Ni$_{80}$Cr$_{20}$) contacts are present. Due to the opposite Seebeck coefficients of Permalloy (S=-20$\mu$V/K) and NiChrome (S=20$\mu$V/K) both thermocouples (contact 4-5 and 6-7) have a thermal sensitivity of S$_{Py-NiCr}$ $\approx$ 40 $\mu$V/K and effectively measure the temperature of the magnet under the Nichrome contacts.

\begin{figure}[t]
\includegraphics[width=8.8cm,keepaspectratio=true]{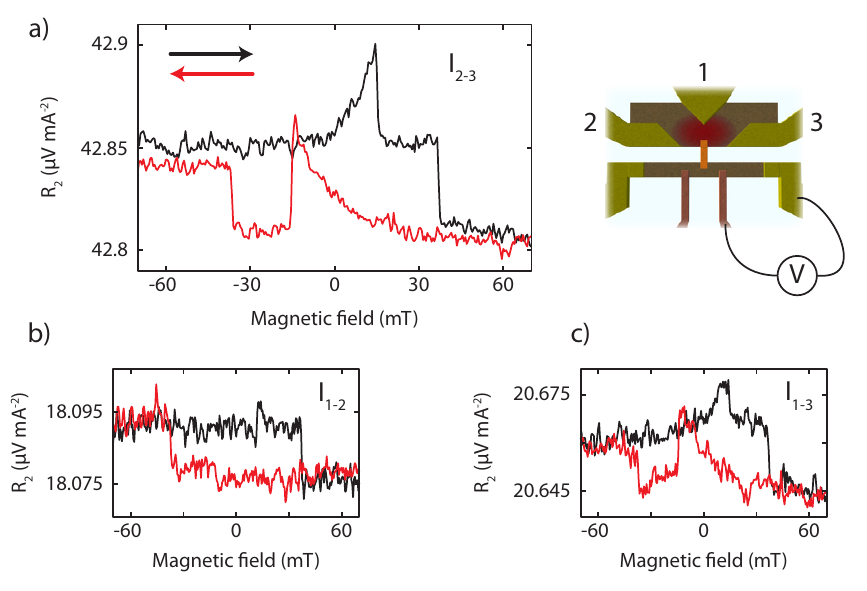}
\caption{\label{fig:3} (color online) Measured voltage from the Py-NiCr thermocouple by selectively switching the magnetizations by an antiparallel magnetic field in a lateral spin valve. The right thermocouple is measured when the Joule heating current is sent a) parallel to the magnetization M$_{1}$, b) under a 45 degree angle and c) under a -45 degree angle. The results were calculated by sending an r.m.s charge current of 1.5 mA. The other thermocouple was also measured and shows similar results.}
\end{figure}

The device was fabricated in a 1 step optical and 6 step electron beam lithography. First, large 150/5 nm thick Ti/Au contacts are made using an optical lithography step and electron beam deposition after which 100 nm wide and 30/5 nm thick Ti/Au markers are fabricated using electron beam lithography. In the subsequent lithography steps, 15 nm thick Permalloy, a 30/5 nm thick Ti/Au interlayer, 5/170 nm thick Ti/Au, 45 nm NiCr and 60 nm Copper were deposited using electron beam deposition.

In our experiment, we selectively switch the magnetizations of both magnets FM$_{1}$ and FM$_{2}$ by applying an antiparallel magnetic field and observe the heat transported through the spin valve by Joule heating FM$_{1}$ and measuring the voltage of the thermocouples on FM$_{2}$. Since the Joule heating scales with I$^{2}$ we are only interested in the R$_{2}$ ($\mu$V/mA$^{2}$) component of the measured voltage V=R$_{1}$I+R$_{2}$I$^{2}$\ldots which we determine by performing lock-in measurements\cite{Bakker,Slachter}. All measurements were done at room temperature.

How exactly the anomalous-Nernst effect and anisotropic magnetoresistive heating can be measured in this device is illustrated in figure 2. The generated heat in the device is transported by the four contacts making up the two thermocouples. At the NiCr contacts the heat is transported in the plane of the device while at the gold contacts this predominantly takes place perpendicular to the plane of the device owing to the difference in thermal conductivity between the materials. Since the magnetization of FM$_{2}$ points along the easy axis of the magnet the anomalous-Nernst effect generates a small voltage difference between both contacts. The sign of this voltage difference changes when the magnetization direction M$_{2}$ flips.

In the same device there are three contacts connected to FM$_{1}$ to send the current either aligned parallel to the magnetization direction (I$_{2-3}$) or under a $\pm$45 degree angle (I$_{1-2}$ or I$_{1-3}$). When the opposing magnetic field in a spin valve has a small angle with respect to the antiparallel direction of the magnetization M$_{1}$, the magnetization rotates prior to the switching process which either increases or decrease the Joule heating. This effect should be pronounced when the current is send under a $\pm$45 degree angle as the change in conductance is then linearly dependent on this deviation angle of the magnetization with the easy axis while in the parallel case this depends quadratically on this angle.

\begin{figure}[!t!]
\includegraphics[width=8.8cm,keepaspectratio=true]{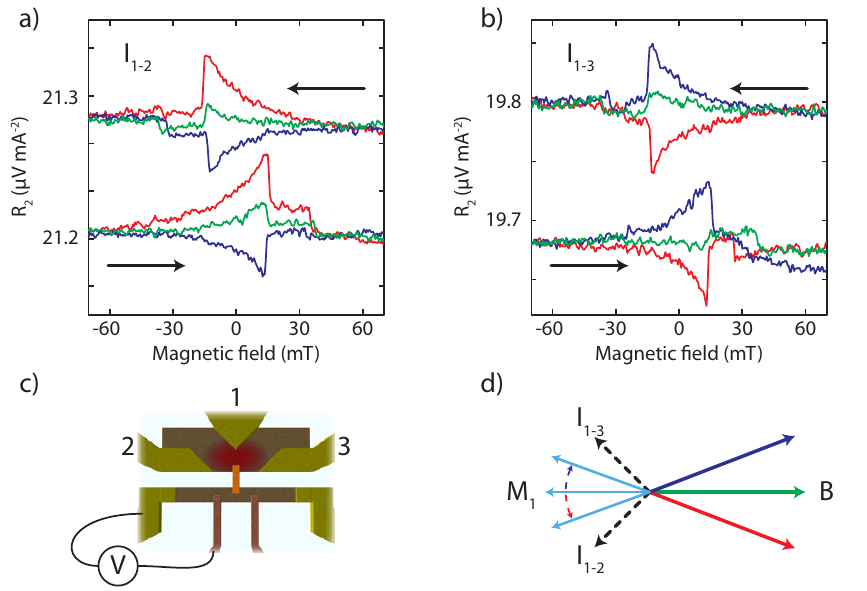}
\caption{\label{fig:4} (color) Measured voltage from the Py-NiCr thermocouple by selectively switching the magnetizations by an opposing magnetic field at 0 and $\pm$10 degrees. The left thermocouple is measured when a) sending the current under a 45 degree angle and b) under a -45 degree angle with respect to the easy magnetization axis. c) The measured configuration and d) the magnetic field configuration and current direction for the measurements of a) \& b). The other thermocouple was also measured and shows similar results.}
\end{figure}

The measured nonlinear voltage R$_{2}$($\mu$V/mA$^{2}$) from the thermocouples is shown for different orientations of the currents in figure 3. Owing to the different dimensions of the ferromagnets, FM$_{1}$ switches by an antiparallel magnetic field of approximately 15 mT while FM$_{2}$ switches at approximately 40 mT. We observe a clear change in the voltage at the switching field of FM$_{2}$. We see this voltage depends only on the orientation of the magnetization of FM$_{2}$. Owing to the finite field at which this magnetization changes sign, the measurement shows a hysteresis loop. When the current is sent parallel to the magnetization, a voltage of 37 nV/mA$^{2}$ can be measured depending on the orientation of M$_{2}$ on top of a large 42.83 $\mu$V/mA$^{2}$ background originating from the temperature measured by the Py-NiCr thermocouple. When the current is sent under a 45 degree angle we measure a smaller 18 nV/mA$^{2}$ signal on top of a smaller 18.085 and 20.65 $\mu$V/mA$^{2}$ background owing to the smaller current path which reduces the Joule heating. We note that the switches do not depend on the thermocouple we measured.

In addition we see a feature appearing prior to the switching of FM$_{1}$ which is different in size depending on the current direction. We believe this can be attributed to anisotropic magnetoresistive heating. To confirm this, we performed our measurements using a magnetic field 10 degree clockwise or anticlockwise to the antiparallel of the magnetizations for the $\pm$45 degree angles between the current we sent through FM$_{1}$ and the magnetization axis. The results of these measurements are shown in figure 4.

\begin{figure}[!t!]
\includegraphics[width=8.8cm,keepaspectratio=true]{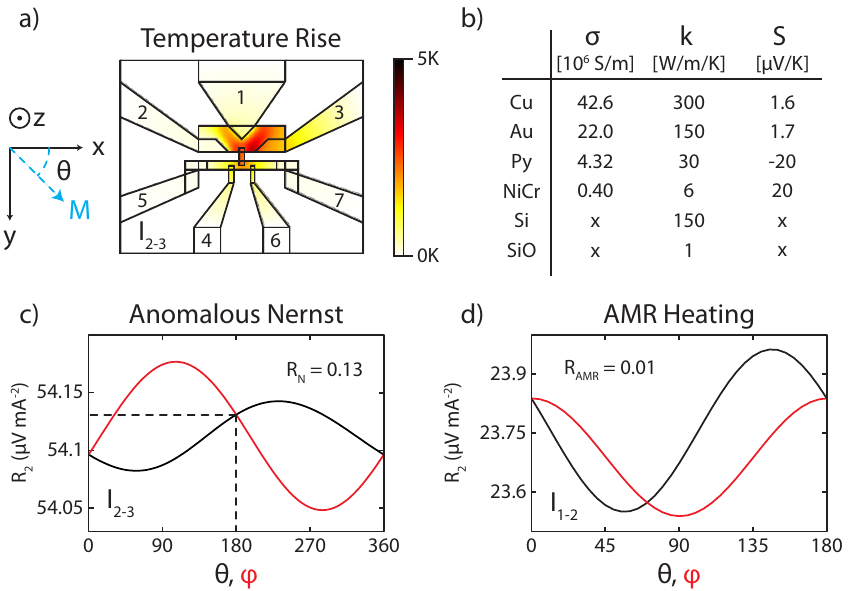}
\caption{\label{fig:5} (color online) Simulated results of the three dimensional thermoelectric model. a) The temperature distribution of the device with a current of 1 mA sent parallel (I$_{2-3}$) to the alignment of the magnetization. b) Input parameters used for the model. The electrical conductivities $\sigma$ are measured while the others are taken from literature\cite{Kittel,Thompson}. c) The simulated anomalous-Nernst voltage from the thermocouples 4-5 and 6-7 as a function of the magnetization angles $\theta$ on the x-y axis and $\phi$ between the x-y plane and the z axis of FM$_{2}$. d) The simulated anisotropic Magnetoresistive heating as a function of the magnetization direction of FM$_{1}$.}
\end{figure}

We clearly observe that the anisotropic magnetoresistive heating increases or decreases by rotating the magnetization prior to switching and has the correct symmetry for a ferromagnetic resistance which is higher for the parallel alignment of the magnetization and current. The voltages arising from this effect are up to 50 nV/mA$^{2}$ in magnitude on top of a 19.7 and 21.1 $\mu$V/mA$^{2}$ background showing that this effect increases or decreases the heating measured by the thermocouple by approximately 0.25\%. The small remaining feature appearing at 0 degrees for the experiment in figure 3 is attributed to the non perfect alignment of the magnetic field in our experiment. We also note that owing to the high conductivity of gold contact 1, the current path does not go exactly straight through the ferromagnet when the current is sent from contact 2 to contact 3. The current path is slightly short circuited which leads to a significant component of the current path which is non-collinear to the magnetization. This effect can be seen by the strong anisotropic magnetoresistive heating component of figure 3a).

In order to quantify the size of the anomalous-Nernst effect and anisotropic magnetoresistive heating, we extend the thermoelectric model used in ref. \cite{Bakker} to include these effects. We use a set of differential equations given by the conservation of charge and heat currents:

\begin{equation}\label{eq:TEmodel}
\left( \begin{array}{c} \vec J \\ \vec Q \end{array} \right) = -\left( \begin{array}{cc} \sigma & \sigma S \\ \sigma \Pi & k \end{array} \right) \left( \begin{array}{c} \vec\nabla V \\ \vec\nabla T \end{array}\right)
\end{equation}

\noindent where $\vec J$ and $\vec Q$ are the charge and heat currents which are related to the voltage gradient $\vec \nabla$V and temperature gradient $\vec \nabla$T by the electrical conductivity $\sigma$, thermal conductivity k, Seebeck coefficient S and Peltier coefficient $\Pi$=S T$_{0}$ with T$_{0}$=293.15 K the reference temperature of the device. The conservation of these currents is given by $\vec \nabla$J = 0 and $\vec \nabla$Q = J$^{2}$/$\sigma$, where we have included Joule heating. The model introduced in ref. \cite{Bakker} is an isotropic model with isotropic coefficients $\sigma$, k and S. We include anisotropic magnetoresistance and the anomalous-Nernst effect by adding anisotropic components to $\sigma$ and S respectively.

Anisotropic magnetoresistance for the magnetization pointing in the direction of any of the three principle axis can be included by using a diagonal 3x3 conductivity matrix $\bf{\sigma}$ with $\sigma_{\parallel}$ on one element of the diagonal and $\sigma_{\perp}$ on the other elements. When the magnetization points in a arbitrary direction given by the angles $\theta$ and $\phi$ this diagonal matrix rotates by $\bf{R \sigma R^{-1}}$ where $\bf{R}$ is the rotation matrix which rotates the ($\parallel,\perp_{1},\perp_{2}$) axes to the (x,y,z) axes. This matrix then becomes:

\begin{equation}\label{eq:TransverseSeebeck}
\sigma_{ij}=\sigma_{\perp}\left(\delta_{ij}-R_{AMR} m_{i} m_{j} \right)
\end{equation}

\noindent where i,j=x,y,z, m$_{i}$ are the x,y,z components of the unit vector $\vec m$ pointing in the direction of the magnetization and $\delta_{ij}$ is the Kronecker delta.

We include the anomalous-Nernst effect by including equation \ref{eq:AnNernst} into the currents defined in equation \ref{eq:TEmodel}. The Seebeck coefficient S now becomes a skew symmetric matrix $\mathbf{S}$ given by:

\begin{equation}\label{eq:TransverseSeebeck}
\mathbf{S}_{ij}=S\left(\delta_{ij}-R_{N} \sum_{k} \varepsilon_{ijk} m_{k} \right)
\end{equation}

\noindent where $\varepsilon_{ijk}$ is the Levi-Civita symbol. A top view of the three dimensional geometry used for the finite element model is shown in figure 5. We included a piece of 2.2 x 3 $\mu$m of the device and set the temperature at all electrical contacts to T$_{0}$. All other outer contact areas are electrically and thermally isolating while on the inner contacts we take the heat and charge current continuous. A charge current is sent through the device by putting a charge current boundary condition on contact 2 and the voltage V=0 on contact 1 or 3. The parameters in figure 5b were used to calculate the temperature rise of the device and subsequent voltage measured by the thermocouples. The 300 nm thick Siliconoxide substrate is also modeled, as well as 700 nm of highly thermally conductive n-doped Silicon\cite{Thompson}. The model was calculated at a current of $\pm$1 mA such that we can distinguish the R$_{1}$ and R$_{2}$ response\cite{Bakker,Slachter}.

We first excluded anisotropic magnetoresistance and the anomalous-Nernst effect in our model and calculated the voltages arising at our thermocouples. We calculate this for the measurement geometries shown in figure 3 a), b) and c). A background of respectively R$_{2}$=54.11 $\mu$V/mA$^{2}$, R$_{2}$=23.99 $\mu$V/mA$^{2}$ and R$_{2}$=26.07 $\mu$V/mA$^{2}$ was calculated for these 3 geometries which is around 25\% higher then observed. This small discrepancy is attributed to the precision of the parameters used.

In the following, we calculate the contribution from the anomalous-Nernst effect to this background voltage. We focus on the measurement geometry and result given in figure 3 a). Figure 5 b) shows the calculated voltage as a function of the FM$_{2}$ magnetization angles $\theta$ in the x-y plane and $\phi$ perpendicular to the plane. We find that an anomalous-Nernst coefficient of R$_{N}$ = 0.13 accurately predicts the 37 nV voltage observed depending on the magnetization direction pointing along the easy axis of FM$_{2}$. The size of the anomalous-Nernst effect is most sensitive to the out of plane angle $\phi$ because the heat currents are predominantly pointing in the plane of the device. Nevertheless, a finite voltage is expected which is around $\frac{1}{3}$ of the maximum effect calculated for an out of plane magnetization at $\phi$ = 105$^{\circ}$ and $\phi$ = 285$^{\circ}$. Using the Seebeck coefficient of Permalloy S$_{Py}$ = -20$\mu$V/K\cite{Saitoh} this leads to a transverse Seebeck coefficient of S$_{N}$ = -2.6 $\mu$V/K.

The size of this coefficient should be equal to that of the anomalous-Hall coefficient when the semiclassical band model applies\cite{Behnia}. This relates these coefficients by the Mott formula for thermoelectricity. We find that it is somewhat larger then the typical anomalous-Hall coefficient of ferromagnetic metals\cite{Nagaosa} of 10$^{-2}$. However, Permalloy is also around 10 times less conductive then the ordinary ferromagnetic metals. When we take this into account, and also the measured size of the anomalous-Hall coefficient of Permalloy\cite{Smit}, we find that our results are in agreement with a semiclassical band model. We note that the anomalous-Nernst effect in our experiments is mathematically equivalent to the combination of a Righi-Leduc effect and the subsequent conversion of the temperature gradient to a voltage gradient and we therefore do not distinguish between them\cite{Slachter3}.

The anisotropic Magnetoresistive heating is calculated for varying angles $\theta$ and $\phi$ of the magnetization of FM$_{1}$ for the measurement geometry used in figure 4. We use an anisotropic Magnetoresistance coefficient R$_{AMR}$ = 0.01 determined from previous experiments\cite{Costachespinpump}. The result is shown in figure 5 d). The calculated voltage from the thermocouple varies by as much as 400 nV/mA$^{2}$ when the magnetization points at $\theta$ = 60$^{\circ}$ or $\theta$ = 145$^{\circ}$. In our experiments we find that when an opposing magnetic field with a $\pm$10 degree with respect to the magnetization axis is applied the voltage prior the switch of the magnetization is $\approx$50 nV. From the calculations we determine that this corresponds to a deviation of the magnetization angle of FM$_{1}$ with the easy axis of 8 degrees.

In conclusion, we have demonstrated how anisotropic Magnetoresistive heating and the anomalous-Nernst effect can be measured in a dedicated caloritronic device. We used a three-dimensional finite-element-model which includes charge and heat transport to model these effects. We extracted an anomalous-Nernst coefficient of R$_{N}$ = 0.13 for Permalloy and found that the magnetization of a Permalloy nanoscale magnet tilts around 7-8 degrees before switching when an opposing magnetic field at a 10 degree angle to the easy axis is applied.

We would like to acknowledge B. Wolfs, S. Bakker and J.G. Holstein for technical assistance. This work was financed by the European EC Contracts IST-033749 'DynaMax', the `Stichting voor Fundamenteel Onderzoek der Materie' (FOM) and NanoNed.

%\bibliography{dissertation}

\end{document}